\documentclass[12pt]{article}
\usepackage{graphics}
\usepackage{graphicx}
\usepackage{color}

\usepackage{amsfonts}

\usepackage{amsmath}
\usepackage{amstext}
\usepackage{amsopn}
\usepackage{amsbsy}
\usepackage{amscd}
\usepackage{amsxtra}
\usepackage{amsthm}
\usepackage{enumerate}

\numberwithin{equation}{section}

\renewcommand{\abstractname}

\title{Multi-scale turbulence modeling and maximum information principle. Part 1}

\author{L. Tao\thanks{ Department of Aerospace Engineering, Indian Institute of Technology Madras, Chennai 600 036, India.
 Email: luoyitao@iitm.ac.in;\ luoyitao@yahoo.com}\ \ and 
M. Ramakrishna\thanks{ Department of Aerospace Engineering, Indian Institute of Technology Madras, Chennai 600 036, India.
Email: krishna@ae.iitm.ac.in} }
\date{}

\begin{document}
\maketitle
\def\s{\!}
\def\ss{\!\!}
\def\sss{\!\!\!}
\def\l{\left}
\def\r{\right}
\def\bx{{\bf x}}
\def\by{{\bf y}}
\def\bz{{\bf z}}
\def\ba{{\bf a}}
\def\bw{{\bf w}}
\def\tbw{\tilde{\bf w}}
\def\bV{{\bf V}}
\def\Vi{V_i}
\def\Vj{V_j}
\def\Vk{V_k}
\def\Vij{\Vi,_j}
\def\Vji{\Vj,_i}
\def\Vkl{\Vk,_l}
\def\bv{{\bf v}}
\def\vi{v_i}
\def\vj{v_j}
\def\tw{\tilde{w}}
\def\wi{w_i}
\def\twi{\tilde{w}_i}
\def\wj{w_j}
\def\twj{\tilde{w}_j}
\def\wk{w_k}
\def\twk{\tilde{w}_k}
\def\wl{w_l}
\def\twl{\tilde{w}_l}
\def\bm{{\bf m}}
\def\bmp{\bm^{\prime}}
\def\bmpp{\bm^{\prime\prime}}
\def\mi{m_i}
\def\bbm{(\bm)}
\def\bbmp{\s\l(\bmp\r)}
\def\bM{{\bf M}}
\def\bbM{\s\l(\bM;\bx\r)}
\def\bn{{\bf n}}
\def\bnp{\bn^{\prime}}
\def\bnpp{\bn^{\prime\prime}}
\def\bbn{(\bn)}
\def\bbnp{\s\l(\bnp\r)}
\def\bk{{\bf k}}
\def\bkp{\bk^{\prime}}
\def\bbk{(\bk)}
\def\bbkp{\s\l(\bkp\r)}
\def\bK{{\bf K}}
\def\bbK{\s\l(\bK;\bx\r)}
\def\bl{{\bf l}}
\def\blp{\bl^{\prime}}
\def\bbl{(\bl)}
\def\bblp{\s\l(\blp\r)}
\def\bi{{\bf i}}
\def\bip{\bi^{\prime}}
\def\bbi{\s\l(\bi\r)}
\def\bbip{\s\l(\bip\r)}
\def\bj{{\bf j}}
\def\bjp{\bj^{\prime}}
\def\bbj{(\bj)}
\def\bbjp{\s\l(\bjp\r)}
\def\bL{{\bf L}}
\def\bp{{\bf p}}
\def\bpp{\bp^{\prime}}
\def\bbp{(\bp)}
\def\bbpp{(\bpp)}
\def\bq{{\bf q}}
\def\bqp{\bq^{\prime}}
\def\p{p}
\def\bbq{(\bq)}
\def\bbqp{\s\l(\bqp\r)}
\def\P{P}
\def\q{q}
\def\tq{\tilde{q}}
\def\barwiwj{\overline{\wi \wj}}
\def\barwiwjwk{\overline{\wi \wj \wk}}
\def\barwixwjy{\overline{\wi(\bx,t)\, \wj(\by,t)}}
\def\barwixwjywkz{\overline{\wi(\bx,t)\,\wj(\by,t)\,\wk(\bz,t)}}
\def\barwixwjywkzwla{\overline{\wi(\bx,t)\,\wj(\by,t)\,\wk(\bz,t)\,\wl(\ba,t)}}
\def\barbwbw{\overline{\bw\bw}}
\def\bartwimtwjn{\overline{\twi\bbm\twj\bbn}}
\def\bartwimtwln{\overline{\twi\bbm\twl\bbn}}
\def\bartwiktwjl{\overline{\twi\bbk\,\twj\bbl}}
\def\bartwititwjtj{\overline{\twi(t;\bi)\twj(t;\bj)}}
\def\bartwitktwjtl{\overline{\twi(t;\bk)\twj(t;\bl)}}
\def\bartwititwjtjtwktk{\overline{\twi(t;\bi)\,\twj(t;\bj)\,\twk(t;\bk)}}
\def\B{{\cal B}}
\def\DD{{\cal D}}
\def\DDi{\DD\s\l(i\r)}
\def\DDj{\DD\s\l(j\r)}
\def\DDk{\DD\s\l(k\r)}
\def\n{{\rm N}}
\def\nO{{\cal O}\s\l(\n\r)}
\def\b{b}
\def\Aj{A_j}
\def\Al{A_l}
\def\Ak{A_k}
\def\Bi{B_i}
\def\Bj{B_j}
\def\Bk{B_k}
\def\Bl{B_l}
\def\Bm{B_m}
\def\Cji{C_{ji}}
\def\Cli{C_{li}}
\def\Clk{C_{lk}}
\def\Cjk{C_{jk}}
\def\Cjl{C_{jl}}
\def\Clm{C_{lm}}
\def\Cnm{C_{nm}}
\def\D{D}
\def\E{E}
\def\EE{{\cal E}}
\def\H{H}
\def\Hij{\H_{ij}}
\def\Wi{W_i}
\def\Wk{W_k}
\def\Wl{W_l}
\def\WiHOMS{\overline{W}_i}
\def\WkHOMS{\overline{W}_k}
\def\WlHOMS{\overline{W}_l}
\def\W{W}
\def\WHOMS{\overline{\W}}
\def\Wij{W_{ij}}
\def\Wil{W_{il}}
\def\Wjl{W_{jl}}
\def\Wkl{W_{kl}}
\def\WijHOMS{\overline{W}_{ij}}
\def\WilHOMS{\overline{W}_{il}}
\def\WjlHOMS{\overline{W}_{jl}}
\def\WklHOMS{\overline{W}_{kl}}
\def\soc{\beta}
\def\socS{\Pi}
\def\toc{\gamma}
\def\foc{\delta}
\def\foocG{\delta^G}
\def\siocG{\zeta^G}
\def\eiocG{\theta^G}
\def\bN{{\bf N}}
\def\bH{{\bf H}}
\def\bR{\mathbb{R}}
\def\bX{{\bf X}}
\def\barbwbwbw{\overline{\bw\bw\bw}}
\def\G{{\cal G}}
\def\xi{x_i}
\def\xj{x_j}
\def\xk{x_k}
\def\xl{x_l}
\def\balpha{{{\bf \alpha}}}
\def\mbi{b_i}
\def\mbj{b_j}
\def\mbk{b_k}
\def\inform{I}
\def\pdf{f}
\def\pdfG{f_G}
\def\pdfD{f_D}
\def\pdfL{f^{(L)}}
\def\pdfH{f^{(H)}}
\def\varvec{\hat{\hat{\bw}}}
\def\Reprevarvec{\hat{\hat{\bw}}}
\def\ReprevarvecL{\hat{\hat{\bw}}^{(L)}}
\def\ReprevarvecH{\hat{\hat{\bw}}^{(H)}}
\def\bkH{{\bf k}^{(H)}}
\def\LagMultiplier{\lambda}
\def\bK{{\bf K}}
\def\hhw{\hat{\hat{w}}}
\def\Det{\inform_D}
\def\K{\inform_T}
\def\vort{\omega}
\def\BETA{\pmb{\beta}}
\def\ip{i^{\prime}}
\def\jp{j^{\prime}}
\def\kp{k^{\prime}}
\def\lp{l^{\prime}}
\def\twip{\tilde{w}_{\ip}}
\def\twjp{\tilde{w}_{\jp}}
\def\twkp{\tilde{w}_{\kp}}
\def\twlp{\tilde{w}_{\lp}}

\begin{abstract}
We discuss averaged turbulence modeling of multi-scales of length for an incompressible Newtonian fluid, with the help of the maximum information principle. 
 We suppose that there exists a function basis to decompose the turbulent fluctuations in a flow of our concern into the components associated with various spatial scales and that there is a probability density function $\pdf$ of these fluctuation components. 
 The unbiased form for $\pdf$ is determined and the turbulence model is closed, with the multi-scale correlations up to the fourth order, through maximizing the information under the constraints of equality and inequality for that flow.
 Due to the computational difficulty to maximize the information, a closely related but simple alternative objective is sought, like the determinant or the trace of the second order correlations of the turbulent flow. Some preliminary results and implications from the application to homogeneous turbulence are presented. 
 Some issues yet to be resolved are indicated.
\end{abstract}

\section{\label{sec:Introduction}Introduction}
 \ \ \ \ There are two widely used methodologies in Reynolds averaged turbulence modeling: One adopts multi-point multi-scale correlations of fluctuations, while limited to homogeneous turbulence (cf.~\cite{Davidson2004} and \cite{SagautCambon2008}); The other, without the restriction of homogeneous turbulence, adopts one-point correlations and builds widely used engineering models (cf.~\cite{Launder1996} and \cite{Pope2000}). There are also developments of multi-point or mixed closure models for inhomogeneous turbulence (cf.~\cite{Cambon2002} and \cite{Schiestel2008}). Exploring a different strategy, Tao, et al. (cf.~\cite{TaoRamakrishnaRajagopal2008}) have attempted to unify the methodologies by both employing multi-scale correlations to inhomogeneous turbulence and resolving the issue of closure through multi-objective optimization under various constraints of equality and inequality. These constraints are either general and intrinsic to turbulence or special to a particular flow concerned.

Specifically, Tao, et al. (cf.~\cite{TaoRamakrishnaRajagopal2008}) have presented a rationale for statistically averaged multi-scale turbulence modeling for an incompressible Newtonian fluid on the basis of optimizing certain objective functions motivated by homogeneous turbulence. From the viewpoint of information availability, they have justified why their formulation is restricted to the resolution of the average flow fields up to the second order correlations and to the modeling of the third and fourth order correlations in algebraic forms. Also, they have explained why they choose an optimization approach to help make the average model determinate, which can be understood from the need to satisfy the constraints as to be listed in Section~\ref{sec:BasicFormulation}. 

Several objective functions are tentatively proposed to be minimized in \cite{TaoRamakrishnaRajagopal2008}. One difficulty encountered in the proposal is the development of a computational scheme to carry out the multi-objective optimization itself and the selection of an optimal solution among the possibly many; the alternative is to seek a single objective function to be optimized as discussed in this work. Another difficulty faced is that there appear no clear-cut rules to guide in the determination of the turbulence model construction. We attempt to resolve these issues in this work, with the help of the maximum information principle (cf.~\cite{Haken2006} and \cite{Jaynes2003}). We should also point out that some objectives proposed in \cite{TaoRamakrishnaRajagopal2008} are incorrect, such as the minimization of the fluctuation kinetic energy.

One major issue in turbulence modeling is the lack of data on higher order correlations, e.g., about the boundary conditions for the one-point correlations higher than the Reynolds stress in an inhomogeneous turbulent flow. This is the issue of information unavailability. One of its implications is that we encounter difficulties to test, calibrate or apply a turbulence model containing higher order correlations as part of its primary fields. Also, a model containing higher order correlations faces the formidable challenge in the aspect of mathematical analysis and numerical computation due to the great number of variables, constraints and evolution equations involved. Therefore, we intend to construct a model including and resolving only the lower order correlations of the velocity and pressure fluctuations, and thus, the model cannot be expected to provide accurate data for the correlations beyond. If we construct such a model, it is informationally sound to have a corresponding probability density function which may produce adequate results for the included correlations with the maximum `spread-out' allowed, i.e., the probability density function is unbiased. The maximum information principle offers such a tool, which is pursued in this work.
 Moreover, the principle apparently leads to the maximum entropy production in the aspect of the micro-states defined by the velocity fluctuations of various spatial scales.
 The principle has single objective function, the information, to be maximized, a preferable alternative to multi-objective optimizations.

We are aware of the controversy and the problem surrounding the maximum information principle, especially in the case of continuous probability distributions.
 We should also mention the question of the applicability of the principle to turbulence model construction, considering the non-equilibrium nature of turbulent motion.
 Here, we intend to explore the possibility and consequence of applying the principle to turbulence modeling and to offer an interesting test of the principle itself.
 We will see that the adoption of the maximum information principle leads to a closed multi-scale turbulence model up to the fourth order correlations.
 There still remains a tremendous task to make the model computationally feasible to turbulence simulation, as to be delineated later. 

For an objective function involving the probability density function explicitly, such as the information in the maximum information principle, there appears an intrinsic difficulty to make the objective optimization computationally feasible which will become clear in Section~\ref{sec:MaximumInformationPrinciple}. This motivates us to seek some closely related but computationally less demanding alternative objective functions to be maximized, like the determinant or the trace of the second order correlations and so on.        

To check whether the present model may produce meaningful results, we apply it to the special case of homogeneous turbulence; some preliminary results are given so as to indicate its potential.

The present work intends to explore a non-traditional methodology to resolve the issue of closure and construction of multi-scale turbulence models.
 It indicates, in details, the challenge to construct an averaged turbulence model resolving the multi-scale correlations of velocity fluctuations and pressure fluctuations, in comparison with engineering turbulence modeling.

This paper is organized as follows.
 In Section~\ref{sec:BasicFormulation}, we present the basic ideas about the turbulence modeling of multi-scales and the constraints of equality and inequality for the fluctuation correlations up to the fourth order.
 We explore in Section~\ref{sec:MaximumInformationPrinciple} the maximum information principle and its consequence in turbulence modeling, with the information as the single objective function to be maximized.
 In Section~\ref{sec:ApproximateObjective}, as a simplification of and an approximation to the information, we discuss the invariants of the second order correlation as single objectives.
 In Section~\ref{sec:HomogeneousTurbulence}, we apply the model to homogeneous turbulence and discuss some preliminary results and implications.
 Some concluding remarks are made in Section~\ref{sec:Remarks} about the present formulation and some of the issues to be investigated.

\section{\label{sec:BasicFormulation}Basic Formulation}
\ \ \ \ Consider the isothermal motion of an incompressible Newtonian fluid in a finite domain $\DD\subset{\bR}^3$ at time $t\geq 0$. The fields for the velocity and pressure are governed by the Navier-Stokes equations, i.e., 
\begin{eqnarray}
\vj,_j=0,
\label{mass}
\end{eqnarray}
\begin{eqnarray}
\frac{\partial\vi}{\partial t}+\l(\vi\vj\r)\s,_j=-\,\p,_i + \,\nu\,\vi,_{jj}.
\label{NS}
\end{eqnarray}
Here  $\nu$ is the kinematic viscosity of the fluid; the mass density $\rho$ is incorporated into $p$. Adopting the standard practice in Reynolds averaging, we take the partition
\begin{eqnarray}
\vi=\Vi+\wi,\ \ 
\p=\P+\q,\ \
\overline{\wi}=0,\ \
\overline{\q}=0,
\label{R_p}
\end{eqnarray}
where $\Vi$ and $\rho\P$ are the ensemble averaged velocity and pressure which behave regularly; $\wi$ and $\rho\q$ are the parts associated with the random fluctuations of the velocity and pressure fields, if the flow is turbulent. Then, (\ref{mass}) through (\ref{R_p}) result in
\begin{eqnarray}
\Vj,_j=0,
\label{mass_avg}
\end{eqnarray}
\begin{eqnarray}
\frac{\partial\Vi}{\partial t}+\l(\Vi\Vj+\barwiwj\r)\s,_j
=-\P,_i+\,\nu\,\Vi,_{jj},
\label{NS_avg}
\end{eqnarray}
\begin{eqnarray}
\wj,_j=0,
\label{mass_fluct}
\end{eqnarray}
\begin{eqnarray}
\frac{\partial\wi}{\partial t}+\l(\wi\wj+\wi\Vj+\Vi\wj-\barwiwj\r)\s,_j
=-\q,_i+\,\nu\,\wi,_{jj},
\label{NS_fluct}
\end{eqnarray}
and

\vskip -12pt
\begin{eqnarray}
-\q,_{jj}=\l(\wi\wj+2\wi\Vj-\barwiwj\r)\s,_{ij}.
\label{pressu_fluct}
\end{eqnarray}

Next, as done in \cite{TaoRamakrishnaRajagopal2008}, we assume that there is an orthonormal basis for the functions defined in $\DD$\footnote{In the case of $\DD$ being a cuboid, we may use the Cartesian coordinate system and the Legendre polynomials or the Chebyshev polynomials, etc. If the turbulence is homogeneous, one may resort conventionally to Fourier series. There are cases in which a curvilinear coordinate system and other function basis may be preferred. For a rather generally shaped $\DD$, we might still employ the Legendre polynomials or the Chebyshev polynomials as the function basis (in the Cartesian coordinate system with a cuboid containing $\DD$), however, the functions defined on $\DD$ need to be extended to the cuboid properly.}
\begin{align}
\B:=\bigg\{\b\s\l(\bx;\bm\r)\s:& \ 
\bx\in\DD,\ \bm=(m_1,m_2,m_3),\ m_k\in\{0\}\cup\mathbb{N},\
\int_{\DD}\b\s\l(\bx;\bm\r)\b\s\l(\bx;\bn\r)d\bx=\delta_{\bm}\,_{\bn}
\bigg\},
\end{align}
so that the fluctuations $\wi$ and $\q$ can be adequately represented as
\begin{eqnarray}
\wi=\sum_{\bm}\twi\s\l(t;\bm\r)\b\s\l(\bx;\bm\r), \quad 
\q =\sum_{\bm}\tq\s\l(t;\bm\r) \b\s\l(\bx;\bm\r)\s .
\label{wqrep}
\end{eqnarray}
Each $\bm$ may be interpreted as a wave number characterizing certain spatial scales, along with its associated $\b\s\l(\bx;\bm\r)$ which is oscillatory in space similar to that of Fourier representation; $\twi\s\l(t;\bm\r)$ and $\rho\,\tq\s\l(t;\bm\r)$ are the fluctuation components of velocity and pressure associated with the scale of $\bm$. From the consideration of the deterministic continuum treatment and the viscous dissipation effect, there is expectedly an upper bound on $\bm$, and thus, the number of the terms in the above representations is finite. It is easy to extend the function basis to the case in which a weight is required to define orthogonality. It follows from (\ref{mass_fluct}) through (\ref{wqrep}) that the evolutions of the second and third order correlations of 
\begin{align}
\soc_{ij}\s\l(\bi,\bj\r):=\bartwititwjtj,\quad 
\toc_{ijk}\s\l(\bi,\bj,\bk\r):=\bartwititwjtjtwktk
\end{align}
as well as the correlations of pressure fluctuations are governed by 
\begin{align}
&
\frac{d}{dt}\soc_{ij}\s\l(\bi,\bj\r)
-\nu\sum_{\bm}\D\s\l(\bm;\bi\r)\soc_{ij}\s\l(\bm,\bj\r)
+\sum_{\bm}\Big(\Wil\s\l(\bm,\bi\r)+\W\s\l(\bm;\bi\r)\delta_{il}\Big)
\soc_{lj}\s\l(\bm,\bj\r)
\notag\\
&
-\nu\sum_{\bm}\D\s\l(\bm;\bj\r)\soc_{ji}\s\l(\bm,\bi\r)
+\sum_{\bm}\Big(\Wjl\s\l(\bm,\bj\r)+\W\s\l(\bm;\bj\r)\delta_{jl}\Big)
\soc_{li}\s\l(\bm,\bi\r)
\notag\\
=&
-\sum_{\bm,\bn}\Al\s\l(\bm;\bn,\bi\r)\toc_{lij}\s\l(\bn,\bm,\bj\r)
-\,\sum_{\bm}\Bi\s\l(\bm;\bi\r)\overline{\tq\bbm\twj\bbj}
\notag\\ 
&
-\sum_{\bm,\bn}\Al\s\l(\bm;\bn,\bj\r)\toc_{lji}\s\l(\bn,\bm,\bi\r)
-\sum_{\bm}\Bj\s\l(\bm;\bj\r)\overline{\tq\bbm\twi\bbi},
\label{dtwitwjdt}
\end{align}
\begin{align}
-\sum_{\bm}\D\s\l(\bm;\bp\r)\overline{\tq\bbm\twi\bbi}=2\sum_{\bm}\Wk\s\l(\bm;\bp\r)\soc_{ki}\s\l(\bm,\bi\r)+
\sum_{\bm,\bn}\Clk\s\l(\bm,\bn;\bp\r)\toc_{kli}\s\l(\bm,\bn,\bi\r),
\label{dtqtwidt}
\end{align}
\begin{align}
&
\frac{d}{dt}\toc_{ijk}\s\l(\bi,\bj,\bk\r)
-\nu\sum_{\bm}\D\s\l(\bm;\bi\r)\toc_{ijk}\s\l(\bm,\bj,\bk\r)
+\sum_{\bm}\Big(\Wil\s\l(\bm,\bi\r)+\W\s\l(\bm;\bi\r)\delta_{il}\Big)
\toc_{ljk}\s\l(\bm,\bj,\bk\r)
\notag\\
&
-\nu\sum_{\bm}\D\s\l(\bm;\bj\r)\toc_{jki}\s\l(\bm,\bk,\bi\r)
+\sum_{\bm}\Big(\Wjl\s\l(\bm,\bj\r)+\W\s\l(\bm;\bj\r)\delta_{jl}\Big)
\toc_{lki}\s\l(\bm,\bk,\bi\r)
\notag\\ 
& 
-\nu\sum_{\bm}\D\s\l(\bm;\bk\r)\toc_{kij}\s\l(\bm,\bi,\bj\r)
+\sum_{\bm}\Big(\Wkl\s\l(\bm,\bk\r)+\W\s\l(\bm;\bk\r)\delta_{kl}\Big)
\toc_{lij}\s\l(\bm,\bi,\bj\r)
\notag\\
=&
-\sum_{\bm,\bn}\Al\s\l(\bm;\bn,\bi\r)\Big(
\foc_{lijk}\s\l(\bn,\bm,\bj,\bk\r)
-\soc_{li}\s\l(\bn,\bm\r)\soc_{jk}\s\l(\bj,\bk\r)\Big)
-\sum_{\bm}\Bi\s\l(\bm;\bi\r)\overline{\tq\bbm\twj\bbj\twk\bbk}
\notag\\
&
-\sum_{\bm,\bn}\Al\s\l(\bm;\bn,\bj\r)\Big(
\foc_{ljki}\s\l(\bn,\bm,\bk,\bi\r)
-\soc_{lj}\s\l(\bn,\bm\r)\soc_{ki}\s\l(\bk,\bi\r)\Big)
-\sum_{\bm}\Bj\s\l(\bm;\bj\r)\overline{\tq\bbm\twk\bbk\twi\bbi}
\notag\\
&
-\sum_{\bm,\bn}\Al\s\l(\bm;\bn,\bk\r)\Big(
\foc_{lkij}\s\l(\bn,\bm,\bi,\bj\r)
-\soc_{lk}\s\l(\bn,\bm\r)\soc_{ij}\s\l(\bi,\bj\r)\Big)
-\sum_{\bm}\Bk\s\l(\bm;\bk\r)\overline{\tq\bbm\twi\bbi\twj\bbj}\,,
\label{dtwitwjtwkdt}
\end{align}
\begin{align}
-\sum_{\bl}\D\s\l(\bl;\bi\r)
\overline{\tq\bbl\twj\bbj\twk\bbk}
=
&\,
2\sum_{\bm}\Wl\s\l(\bm;\bi\r)\toc_{ljk}\s\l(\bm,\bj,\bk\r)
\notag\\
&
+\sum_{\bm,\bn}\Cli\s\l(\bm,\bn;\bi\r)
\Big(
\foc_{iljk}\s\l(\bm,\bn,\bj,\bk\r)
-\soc_{il}\s\l(\bm,\bn\r)\soc_{jk}\s\l(\bj,\bk\r)\s\Big),
\label{dtqtwitwjdt}
\end{align}
and
\begin{align}
-\sum_{\bl}\D\s\l(\bl;\bi\r)
\overline{\tq\bbl\tq\bbp}
=
2\sum_{\bm}\Wl\s\l(\bm;\bi\r)\overline{\tq\bbp\twl\bbm}
+\sum_{\bm,\bn}\Cjl\s\l(\bm,\bn;\bi\r)\overline{\tq\bbp\twl\bbm\twj\bbn}\,.
\label{dtqtqdt}
\end{align}
Here
\begin{align}
\foc_{ijkl}\s\l(\bi,\bj,\bk,\bl\r):=
\overline{\twi\bbi\twj\bbj\twk\bbk\twl\bbl}
\end{align}
is the fourth order correlation, and
\begin{align}
&
\twi\bbm:=\twi\s\l(t;\bm\r),\quad 
\tq\bbm:=\tq\s\l(t;\bm\r),\quad 
\b\bbm:=\b\s\l(\bx;\bm\r);
\notag\\[3pt]
&
\Aj\s\l(\bm;\bn,\bk\r):=\int_{\DD}\l[\b\bbm\r]\s,_j\b\bbn\b\bbk d\bx,
\quad
\Bi\s\l(\bm;\bn\r):=\int_{\DD}\l[\b\bbm\r]\s,_i\b\bbn d\bx,
\notag\\ &
\Cji\s\l(\bm,\bn;\bk\r):=\int_{\DD}\l[\b\bbm\r]\s,_j\l[\b\bbn\r]\s,_i\b\bbk d\bx,
\quad
\D\s\l(\bm;\bk\r):=\int_{\DD}\l[\b\bbm\r]\s,_{jj}\b\bbk d\bx,
\notag\\ &
\W\s\l(\bm;\bk\r):=\int_{\DD}\l[\b\bbm\r]\s,_{j}\Vj\,\b\bbk d\bx,
\quad
\Wi\s\l(\bm;\bk\r):=\int_{\DD}\l[\b\bbm\r]\s,_{j}\Vji\,\b\bbk d\bx,
\notag\\ &
\Wij\s\l(\bm,\bk\r):=\int_{\DD}\b\bbm\b\bbk\Vij d\bx.
\label{Coefficients}
\end{align}

The Reynolds stress tensor in (\ref{NS_avg}) can be represented as
\begin{eqnarray}
\barwiwj=\sum_{\bi,\bj}\soc_{ij}\s\l(\bi,\bj\r)\b\bbi\,\b\bbj.
\label{Reynolds_stress}
\end{eqnarray}
Equations \eqref{mass_avg}, \eqref{NS_avg}, \eqref{dtwitwjdt} through \eqref{Reynolds_stress} form a closed model of multi-scale turbulence, provided that the fourth order correlations are approximated, like in terms of the lower order quantities. From the solutions of this model, we may construct the multi-point correlations such as, $\forall$ $\bx$, $\by$, $\bz$, $\ba \in \DD$,
\begin{align}
&\barwixwjy=\sum_{\bi,\bj}\soc_{ij}\s\l(\bi,\bj\r)\b(\bx;\bi)\, \b(\by;\bj),
\notag\\
&\barwixwjywkz=\sum_{\bi,\bj,\bk}\toc_{ijk}\s\l(\bi,\bj,\bk\r)\b(\bx;\bi)\,\b(\by;\bj)\,\b(\bz;\bk),
\notag\\
&\barwixwjywkzwla
=\sum_{\bi,\bj,\bk,\bl}\foc_{ijkl}\s\l(\bi,\bj,\bk,\bl\r)\b(\bx;\bi)\,\b(\by;\bj)\,\b(\bz;\bk)\,\b(\ba;\bl),
\notag\\
&\overline{\q\s\l(\bx,t\r) \q\s\l(\by,t\r)}
=\sum_{\bk,\bl}\overline{\tq\s\l(\bk\r)\tq\s\l(\bl\r)}\,\b(\bx;\bk)\,\b(\by;\bl).
\label{MultiPointCorrelations}
\end{align}

The inclusion of \eqref{dtwitwjtwkdt} for the evolution of $\toc_{ijk}\s\l(\bi,\bj,\bk\r)$ is based on the consideration that $\toc_{ijk}\s\l(\bi,\bj,\bk\r)$ and $\foc_{ijkl}\s\l(\bi,\bj,\bk,\bl\r)$ are needed to obtain the correlation of pressure fluctuations as seen above. If we model $\foc_{ijkl}\s\l(\bi,\bj,\bk,\bl\r)$, it is preferable to have an equation of evolution to describe $\toc_{ijk}\s\l(\bi,\bj,\bk\r)$; the treatment preserves the intrinsic relationship between $\toc_{ijk}\s\l(\bi,\bj,\bk\r)$ and $\foc_{ijkl}\s\l(\bi,\bj,\bk,\bl\r)$ as displayed in \eqref{dtwitwjtwkdt} and \eqref{dtqtwitwjdt} and the constraints to be discussed below, and also, it makes the fourth order correlation the sole quantities to be modeled.

This discussion justifies the adoption of a multi-scale turbulence model up to the fourth order correlation. Models involving higher order correlations are not pursued due to the information unavailability and the difficulty of mathematical analysis and computation.
 These problems are encountered even in the present model, especially regarding the inclusion of \eqref{dtwitwjtwkdt}: 
 The first is how to furnish adequate initial conditions for $\toc_{ijk}\s\l(\bi,\bj,\bk\r)$ in the case that the average flow fields are time-dependent -- its resolution may be case dependent. The second is how to furnish boundary conditions for $\overline{\wi\wj\wk}$, in the case that we do not have such data -- we leave their determination as part of the solution from the model through the optimization to be introduced below.

The question now is how to determine $\foc_{ijkl}\s\l(\bi,\bj,\bk,\bl\r)$ such that the model composed of \eqref{mass_avg}, \eqref{NS_avg}, \eqref{dtwitwjdt} through \eqref{Reynolds_stress} is closed. 
 In the following, we explore specifically the scheme of the maximum information principle so as to construct a determinate model through maximizing one objective function called the information. 
  To this end, we discuss first the constraints of equality and inequality on $\soc_{ij}\s\l(\bi,\bj\r)$, $\toc_{ijk}\s\l(\bi,\bj,\bk\r)$ and $\foc_{ijkl}\s\l(\bi,\bj,\bk,\bl\r)$ as an integral part of the formulation of closure and maximization.
\begin{enumerate}
  \renewcommand{\theenumi}{\roman{enumi}}
\item
The definitions of the correlations imply that
\begin{align}
&
\soc_{ij}\s\l(\bi,\bj\r)=\soc_{ji}\s\l(\bj,\bi\r),\quad
\toc_{ijk}\s\l(\bi,\bj,\bk\r)=\toc_{jik}\s\l(\bj,\bi,\bk\r)=\toc_{kji}\s\l(\bk,\bj,\bi\r)=...,
\notag\\[4pt]
&
\foc_{ijkl}\s\l(\bi,\bj,\bk,\bl\r)
=\foc_{jikl}\s\l(\bj,\bi,\bk,\bl\r)
=\foc_{kjil}\s\l(\bk,\bj,\bi,\bl\r)
=\foc_{ljki}\s\l(\bl,\bj,\bk,\bi\r)
=...
\label{Constraints_Symmetry}
\end{align}
That is, each correlation remains invariant under the interchange of any two pairs of its indexes, say $\{i,\bi\}$ and $\{j,\bj\}$ as indicated above.

\item
 Applying the Cauchy-Schwarz inequality to $\soc_{ij}\s\l(\bi,\bj\r)$, $\toc_{ijk}\s\l(\bi,\bj,\bk\r)$ and $\foc_{ijkl}\s\l(\bi,\bj,\bk,\bl\r)$ gives, $\forall i$, $j$, $k$, $l$, $\forall \bi$, $\bj$, $\bk$, $\bl$, (the summation rule suspended),
\begin{align}
&
\soc_{ii}\s\l(\bi,\bi\r)\geq 0,\quad
\soc_{ii}\s\l(\bi,\bi\r)\soc_{jj}\s\l(\bj,\bj\r)-
\big(\soc_{ij}\s\l(\bi,\bj\r)\s\big)^2\geq 0,
\notag\\
&
\foc_{iijj}\s\l(\bi,\bi,\bj,\bj\r)\geq 0,\quad
\soc_{ii}\s\l(\bi,\bi\r)\foc_{jjkk}\s\l(\bj,\bj,\bk,\bk\r)-
\big(\toc_{ijk}\s\l(\bi,\bj,\bk\r)\s\big)^2\geq 0,
\notag\\
&
\foc_{iijj}\s\l(\bi,\bi,\bj,\bj\r)\foc_{kkll}\s\l(\bk,\bk,\bl,\bl\r)-
\big(\foc_{ijkl}\s\l(\bi,\bj,\bk,\bl\r)\s\big)^2\geq 0.
\label{Constraints_CauchySchwarzInWaveNumbers}
\end{align}

\item
 The Reynolds stress tensor should be positive semi-definite, i.e.,
\begin{align}
\overline{w_1 w_1}\geq 0,\quad
\overline{w_1 w_1}\,\,\overline{w_2 w_2}-\big(\overline{w_1 w_2}\big)^2\geq 0,\quad
\text{det}\big[\overline{w_i w_j}\big]\geq 0,
\label{Constraints_ReynoldsStressTensor}
\end{align}
in order to guarantee the non-negativity of both the fluctuation energy and the average bound for the velocity fluctuations.

The quantity of $\overline{\wi,_k \wj,_k}$ should be positive semi-definite,
\begin{align}
\overline{w_1,_k w_1,_k}\geq 0,\quad
\overline{w_1,_k w_1,_k}\,\,\overline{w_2,_l w_2,_l}-\big(\overline{w_1,_k w_2,_k}\big)^2\geq 0,\quad
\text{det}\big[\overline{w_i,_k w_j,_k}\big]\geq 0,
\label{Constraints_ViscousDissipation}
\end{align}
due to its significance associated with the viscous dissipation in turbulence. In symmetry, one may also enforce the positive semi-definiteness of $\overline{\wk,_i \wk,_j}$ similarly.

Considering the importance of vorticity in turbulence (cf.~\cite{Davidson2004}), one may introduce $\vort_{ij}:=(\wi,_j-\wj,_i)/2$ and formulate the constraints of

\vskip -12pt
\begin{align}
&
\overline{(\vort_{12})^2}\geq 0,\quad
\overline{(\vort_{23})^2}\geq 0,\quad
\overline{(\vort_{31})^2}\geq 0,
\quad
\overline{(\vort_{12})^2}\,\,\overline{(\vort_{23})^2}-\big(\overline{\vort_{12}\vort_{23}}\big)^2\geq 0,
\notag\\
&
\overline{(\vort_{23})^2}\,\,\overline{(\vort_{31})^2}-\big(\overline{\vort_{23}\vort_{31}}\big)^2\geq 0,\quad
\overline{(\vort_{31})^2}\,\,\overline{(\vort_{12})^2}-\big(\overline{\vort_{31}\vort_{12}}\big)^2\geq 0,
\label{Constraints_Vorticity}
\end{align}
from the definition and the Cauchy-Schwarz inequality.

It is also natural to require that
\begin{align}
\overline{\q^2}\geq 0,\quad
\overline{\q^2}\,\,\overline{\wi^2}-\l(\overline{\q \wi}\r)^2\geq 0,\quad
\overline{\q^2}\,\,\overline{(\wi\wj)^2}-\l(\overline{\q \wi \wj}\r)^2\geq 0.
\label{Constraints_PressureFluctuation}
\end{align}

\item
 Equations \eqref{mass_fluct} and \eqref{wqrep} result in, $\forall i$, $j$, $k$, $\forall \bi$, $\bj$, $\bk$, $\bn$,
\begin{align}
&
\sum_{\bm}\soc_{im}\s\l(\bi,\bm\r)\Bm\s\l(\bm;\bn\r)=0,
\quad
\sum_{\bm}\toc_{ijm}\s\l(\bi,\bj,\bm\r)\Bm\s\l(\bm;\bn\r)=0,
\notag\\
&
\sum_{\bm}\foc_{ijkm}\s\l(\bi,\bj,\bk,\bm\r)\Bm\s\l(\bm;\bn\r)=0.
\label{Constraints_IsochoricFlow}
\end{align}

\item
We apply the Cauchy-Schwarz inequality to the multi-point correlations of fluctuations in the physical space so as to obtain  the following constraints, i.e., $\forall i$, $j$, $k$, $l$, $m$, $n$, $p$, $\forall \bx$, $\by$, $\bz$, $\ba\in\DD$, (the summation rule suspended),
\begin{align}
&
\overline{\l(\wi(\bx)\r)^2}\,\,\overline{\l(\wj(\by)\r)^2}-
\l(\overline{\wi(\bx)\,\wj(\by)}\r)^2\geq 0,
\quad
\overline{\l(\vort_{ij}(\bx)\r)^2}\,\,\overline{\l(\vort_{kl}(\by)\r)^2}-
\l(\overline{\vort_{ij}(\bx)\,\vort_{kl}(\by)}\r)^2\geq 0,
\notag\\
&
\overline{\l(\wi,_m\s(\bx)\r)^2}\,\,\overline{\l(\wj,_n\s(\by)\r)^2}-
\l(\overline{\wi,_m\s(\bx)\,\wj,_n\s(\by)}\r)^2\geq 0,
\notag\\
&
\overline{\l(\wj(\by)\,\wk(\bz)\r)^2}\geq 0,
\quad
\overline{\l(\wi(\bx)\r)^2} \,\,\overline{\l(\wj(\by)\,\wk(\bz)\r)^2}-
\l(\overline{\wi(\bx)\,\wj(\by)\,\wk(\bz)}\r)^2\geq 0,
\notag\\
&
\overline{\l(\wi(\bx)\,\wj(\by)\r)^2} \,\,\overline{\l(\wk(\bz)\,\wl(\ba)\r)^2}-
\l(\overline{\wi(\bx)\,\wj(\by)\, \wk(\bz)\,\wl(\ba)}\r)^2\geq 0,
\label{Constraints_CauchySchwarzInPhysicalSpace}
\end{align}
and so on, including more spatial derivatives of first order or even higher orders. One may also obtain the constraints involving the pressure fluctuations.

\item
 There are constraints from the boundary conditions for the Reynolds stress tensor on $\partial\DD$, whose forms and number are flow-dependent; there is also the possibility of zero $\overline{\wi\wj\wk}$ and $\overline{\wi\wj\wk\wl}$ on the solid wall boundary of a flow if the wall is smooth and experiences negligible random vibrations. 

If one has additional information about a specific flow, one may formulate it as additional constraints holding in that flow.
\end{enumerate}

Several comments are in order regarding the above constraints.
 (i) The constraints formulated in the physical space reflect partially the link between the equations of evolution \eqref{dtwitwjdt} through \eqref{dtqtqdt} and the counterpart equations of evolution for $\barwixwjy$ and so on in the physical space.
 (ii) The constraints of inequality from the Cauchy-Schwarz inequality are not that tight since the inequality holds for any two integrable functions. As a part of the model, the collective effect of the very many such constraints may be sufficiently severe to the correlations.
 (iii) The constraints of \eqref{Constraints_ReynoldsStressTensor} through \eqref{Constraints_PressureFluctuation} and \eqref{Constraints_CauchySchwarzInPhysicalSpace} hold in $\DD$ and the constraints from the boundary conditions hold on $\partial\DD$, they need to be treated appropriately in order to derive the corresponding constraints for $\soc_{ij}\s\l(\bi,\bj\r)$, $\toc_{ijk}\s\l(\bi,\bj,\bk\r)$ and $\foc_{ijkl}\s\l(\bi,\bj,\bk,\bl\r)$. One possible scheme is to integrate the inequalities (holding in $\DD$) on the domains of $\DD$, $\DD\s\times\s\DD$, etc. with weight functions like $\big(\b\s\l(\bx;\bm\r)\s\big)^2$, $\big(\b\s\l(\bx;\bm\r)\big)^2\,\big(\b\s\l(\by;\bn\r)\s\big)^2$, etc. 
 (iv) The number of the constraints is great, and thus, the approximate treatment of representative wave numbers should be adopted to make the model computationally feasible, as to be discussed in Section~\ref{sec:MaximumInformationPrinciple}. Also, the approximate treatments of locality and local isotropy to be discussed will convert part of equations \eqref{dtwitwjdt} through \eqref{dtqtqdt} into constraints of equality.
 (v) There is also the issue about Kolmogorov's results as well as their modifications (cf.~\cite{Davidson2004} and \cite{Frisch1995}). We need to examine whether these results are the natural outcome of the model with the constraints listed above or the results themselves should be explicitly formulated and imposed as additional information and constraints in the model.

\section{\label{sec:MaximumInformationPrinciple}Maximum Information Principle}
\ \ \ \ To apply the maximum information principle (MIP), we introduce a vector $\varvec$ of components $\l\{ \twi(t;\bk) \r\}$  which are the turbulent velocity fluctuation components of all the spatial scales, at instant of $t$, involved as in \eqref{wqrep}.  We also introduce a probability density function $\pdf$ of the velocity fluctuation components, $\pdf\big(\varvec\big)$. The introduction of this probability density function is justifiable in that we have effectively used it in setting
\begin{align}
&
\int_{{\bf R}^{3N}} \pdf\big(\varvec\big) d\varvec =1,\quad
\int_{{\bf R}^{3N}} \twi\bbi \pdf\big(\varvec\big) d\varvec =\overline{\twi\bbi}=0,
\quad
\int_{{\bf R}^{3N}} \twi\bbi\, \twj\bbj\, \pdf\big(\varvec\big) d\varvec=\soc_{ij}\s\l(\bi,\bj\r),
\notag\\
&
\int_{{\bf R}^{3N}} \twi\bbi\, \twj\bbj\, \twk\bbk\, \pdf\big(\varvec\big) d\varvec=\toc_{ijk}\s\l(\bi,\bj,\bk\r),
\notag\\
&
\int_{{\bf R}^{3N}} \twi\bbi\, \twj\bbj\, \twk\bbk\,\twl\bbl\, \pdf\big(\varvec\big) d\varvec=\foc_{ijkl}\s\l(\bi,\bj,\bk,\bl\r),
\label{PDF_link}
\end{align}
where $N$ is the number of the wave numbers $\{\bm\}$ involved and $3N$ the dimension of $\varvec$.

Additional remarks are worth making about the form of or the variables involved in $\pdf\big(\varvec\big)$. Clearly, this is a partial treatment, in contrast to a formulation of the probability density functional for $\vi$ and $\p$ (cf.~\cite{Beran1968}) and from it the derivations of \eqref{mass_avg}, \eqref{NS_avg}, \eqref{dtwitwjdt}, \eqref{dtqtwidt}, etc. Here, we do not pursue such a mathematically comprehensive and rigorous treatment. We suppose that the turbulent velocity fluctuations can be defined as in \eqref{R_p} from the notion of ensemble average, these fluctuations can be further decomposed into the components associated with various scales of length according to \eqref{wqrep}, and a probabilistic description of these fluctuation components can be carried out. The pressure fluctuations $\tq\s\l(\bm\r)$ can be represented in terms of $\varvec$ as seen from \eqref{pressu_fluct}, \eqref{wqrep}, \eqref{dtqtwidt}, \eqref{dtqtwitwjdt} and \eqref{dtqtqdt}. The effect o
 f the average velocity and pressure on $\pdf$ will be accounted for through certain parameters or coefficients contained in $\pdf$, as to become clear below.  

We may formulate the principle in the form of
\begin{gather}
\inform=-\int_{{\bf R}^{3N}} \pdf\big(\varvec\big)\,\text{ln}\pdf\big(\varvec\big)\,d\varvec
\label{MIP}
\end{gather}
where $\inform$ is the information to be maximized (cf.~\cite{Haken2006} and \cite{Jaynes2003}). Since $\varvec$ may be viewed as representing micro-states of turbulent fluctuations, the maximization of $\inform$ is like the maximum entropy production regarding the micro-states, if all the constraints on the flow are known and imposed.

Now, we maximize \eqref{MIP} under \eqref{PDF_link}, with the help of the method of Lagrange multipliers, to yield
\begin{align}
\pdf\big(\varvec\big)=\exp\s\bigg[
&-\LagMultiplier
-\sum_{\bi} \LagMultiplier_i\bbi \twi\bbi
-\sum_{\bi,\bj} \LagMultiplier_{ij}(\bi,\bj)\, \twi\bbi \twj\bbj 
-\sum_{\bi,\bj,\bk} \LagMultiplier_{ijk}(\bi,\bj,\bk)\, \twi\bbi \twj\bbj \twk\bbk 
\notag\\
& 
-\sum_{\bi,\bj,\bk,\bl} \LagMultiplier_{ijkl}(\bi,\bj,\bk,\bl)\, \twi\bbi \twj\bbj \twk\bbk \twl\bbl
\bigg].
\label{PDFfromMIP}
\end{align}
Here, $\LagMultiplier$, $\LagMultiplier_i\bbi$'s, $\LagMultiplier_{ij}(\bi,\bj)$'s, $\LagMultiplier_{ijk}(\bi,\bj,\bk)$'s and $\LagMultiplier_{ijkl}(\bi,\bj,\bk,\bl)$'s are the Lagrange multipliers enforcing, respectively, the five sets of constraints in \eqref{PDF_link}. 

All the fourth order correlations are yet to be fixed and expected to be determined with the help of the maximum information principle through maximizing $\inform$. Substituting \eqref{PDFfromMIP} into \eqref{MIP}, we have
\begin{align}
\inform=
\,\LagMultiplier
+\sum_{\bi,\bj} \LagMultiplier_{ij}(\bi,\bj)\, \soc_{ij}(\bi,\bj) 
+\sum_{\bi,\bj,\bk} \LagMultiplier_{ijk}(\bi,\bj,\bk)\,\toc_{ijk}(\bi,\bj,\bk) 
+\sum_{\bi,\bj,\bk,\bl} \LagMultiplier_{ijkl}(\bi,\bj,\bk,\bl)\,\foc_{ijkl}(\bi,\bj,\bk,\bl).
\label{MIP_detail_1}
\end{align}
The multipliers of $\LagMultiplier$'s and the correlations up to the fourth order are linked through \eqref{PDF_link}. Theoretically, we may search and determine the values of $\LagMultiplier_{ijkl}(\bi,\bj,\bk,\bl)$'s by maximizing $\inform$. Specifically, we find these values, for a specific flow of our interest, by maximizing $\inform$ under \eqref{mass_avg}, \eqref{NS_avg}, \eqref{dtwitwjdt} through \eqref{Reynolds_stress} and \eqref{PDF_link}, the aforementioned constraints of equality and inequality, the initial conditions for the second and third order correlations, and the initial and boundary conditions for the average velocity as well as the boundary condition for the average pressure. That is, we may solve the turbulent flow of interest. 

We notice that the number of the correlations, the multipliers, the equations and the constraints is of astronomical order of magnitude, and in practice, we have to reduce it significantly with the help of certain approximations physically plausible.
 Also, the maximum information principle is conceptually based on inference in the face of information incompleteness, and more data from the physical consideration will help to narrow the spread-out of $\pdf$ and to improve the turbulence modeling.  
 There are several conventional approximations we may adopt to these ends:
\begin{enumerate}
\item[(a)]
Instead of employing a complete function basis, we will thin the function basis through selecting representative wave numbers and the associated basis functions sparsely and appropriately, and thus, we can reduce the number of the correlations as well as the multipliers to be solved; for instance, we may take only the wave numbers $\bm=(m_1,m_2,m_3)$ whose components $m_1$, $m_2$ and $m_3$ correspond to nearly the same physical length scales.
 This treatment had better be implemented in the representations of equation \eqref{wqrep} so as to derive the equations of evolution for $\soc_{ij}(\bi,\bj)$ and $\toc_{ijk}(\bi,\bj,\bk)$ and the equations for the correlations of pressure fluctuations conveniently and consistently.

\item[(b)]
We resort to the standard notion of locality or cascade such that all the correlations between the wave numbers not neighboring to each other are neglected.  This point needs to be clarified further by quantifying the meaning of neighborhoods. 
 
\item[(c)]
To adopt the conventional treatment of small-scale, locally isotropic turbulence, we define the notion of high wave numbers and treat the velocity fluctuation components associated with these high wave numbers as isotropic. For example, we 
 set $\soc_{ij}(\bi,\bj)=\soc(\bi,\bj)\,\delta_{ij}$, if the associated wave numbers are high.
\end{enumerate}

The approximations above offer some possibilities to reduce the complexity of the closed model, but they do not provide great help to the computation of \eqref{PDF_link} which is formidable due to the high dimensions involved. It is desirable to have a simplified alternative version of \eqref{MIP} which does not involve $\pdf$ explicitly, as to be discussed next.

\section{\label{sec:ApproximateObjective}Approximate Objective Function}
\ \ \ \ It is known that the information $\inform$ for a Gaussian is proportional to the determinant of the covariance matrix $\big[\soc_{ij}(\bi,\bj)\big]$ (cf.~\cite{CoverThomas1991}); Also, from probability theory, the covariance matrix is a major indicator regarding the extent of the spread-out of $\pdf$,
\begin{align}
\soc_{ij}\s\l(\bi,\bj\r)=\int \twi\bbi\, \twj\bbj\, \pdf\big(\varvec\big) d\varvec.
\label{CovarianceMatrix}
\end{align}

The mathematical properties of $\big[\soc_{ij}\s\l(\bi,\bj\r)\big]$ are characterized by its non-negative eigenvalues and equivalently by its invariants; Two invariants of interests are the trace $\sum_{\bk}\soc_{kk}\s\l(\bk,\bk\r)$ and the determinant $\det\s\big[\soc_{ij}\s\l(\bi,\bj\r)\big]$. 
 In the case that the probability density distribution of a turbulent flow is slightly non-Gaussian, the above-mentioned result of \cite{CoverThomas1991} implies that 
\begin{align}
\Det:=\det\s\big[\soc_{ij}\s\l(\bi,\bj\r)\big]
\label{ApproximateObjectiveDet}
\end{align}
 is an adequate objective function, alternative to the information $\inform$. The difficulty of such a choice is the relevant computational complication.
 From the perspective of computational convenience, the use of 
\begin{align}
\K:=\sum_{\bk}\soc_{kk}\s\l(\bk,\bk\r)
\label{ApproximateObjective}
\end{align}
 as an objective is appealing. Another appeal of this choice is its physical significance, considering that the total fluctuation kinetic energy possessed by a turbulent flow in $\DD$ is related to $\K$ through
\begin{gather}
\frac{1}{2} \int_{\DD}\overline{\wk\wk}\,d\bx=\frac{1}{2} \sum_{\bk}\soc_{kk}\s\l(\bk,\bk\r)=\frac{1}{2}\,\K
\end{gather}
(in the case that the function basis is orthonormal). 
 The third appeal is the relatively simple structure of the trace in the space of $\{\hat\foc_{ijkl}\s\l(\bi,\bj,\bk,\bl\r)\}$ defined by  
\begin{align*}
\hat\foc_{ijkl}\s\l(\bi,\bj,\bk,\bl\r):=
\foc_{ijkl}\s\l(\bi,\bj,\bk,\bl\r)-\soc_{ij}\s\l(\bi,\bj\r)\soc_{kl}\s\l(\bk,\bl\r)
\end{align*}
which is motivated by the structures of \eqref{dtwitwjtwkdt} and \eqref{dtqtwitwjdt}.

From the perspective that turbulence arises from instabilities due to the change of some flow parameters in flows which are originally laminar, such as the increase of Reynolds number, a flow tends to have non-trivial turbulent fluctuation energy under proper flow conditions (cf.~\cite{TennekesLumley1972}).
 This does not necessarily imply that $\K$ of \eqref{ApproximateObjective} is the maximum for the flow to occur under the constraints and conditions given; this point may be clear if one takes, say $\Det$ of \eqref{ApproximateObjectiveDet} as the objective function to be maximized.  
 Further analysis and specific simulations have to be done in order to decide which of the invariants is suitable as the alternative objective function  approximating the information $\inform$.

\section{\label{sec:HomogeneousTurbulence}Homogeneous Turbulence}
\ \ \ \ 
To test preliminarily whether the turbulence model constructed above has the potential to produce meaningful results, we apply it to the special case of homogeneous turbulence here. Let us consider a homogeneous turbulent motion in which $\Vi=V_{ij}\,x_j$ in the cube of $\DD$ $=$ $\l[L/2\r.$, $\l.-L/2\r]^3$ with $V_{ij}$ being constant and $V_{kk}=0$. In addition to the constraints of equality and inequality listed in Section~\ref{sec:BasicFormulation}, there are constraints of equality such as
\begin{align}
&
\Big(\overline{\wi(\bx)\,\wj(\bx)}\Big)\s,_m=0,
\quad
\Big(\overline{\wi,_k\s(\bx)\,\wj,_l\s(\bx)}\Big)\s,_m=0,
\quad
\Big(\overline{\wi(\bx)\,\wj(\bx)\,\wk(\bx)}\Big)\s,_m=0,
\notag\\
&
\Big(\overline{\wi(\bx)\,\wj(\bx)\,\wk(\bx)\,\wl(\bx)}\Big)\s,_m=0,
\label{Constraints_HomogeneousTurbulence}
\end{align}
that are special to homogeneous turbulence resulting from its supposed spatial homogeneity. We may integrate these constraints of equality in the domain of $\DD$ with weight functions like $\b\s\l(\bx;\bm\r)$ in order to derive the corresponding constraints of equality for $\soc_{ij}\s\l(\bi,\bj\r)$, $\toc_{ijk}\s\l(\bi,\bj,\bk\r)$ and $\foc_{ijkl}\s\l(\bi,\bj,\bk,\bl\r)$. As a consequence of \eqref{Constraints_HomogeneousTurbulence}, the $\barwiwj$ related terms disappear from \eqref{NS_avg}, \eqref{NS_fluct} and \eqref{pressu_fluct}, and the product $\soc_{ij}\s\l(\bi,\bj\r)$$\,\soc_{kl}\s\l(\bk,\bl\r)$ related terms disappear from \eqref{dtwitwjtwkdt} and \eqref{dtqtwitwjdt}; the average flow field of $\Vi$ and $\P$ is not affected by the correlations. To discuss the consequence of these disappeared terms clearly, we gather all the relevant evolution equations here:  
\begin{align}
&
\frac{d}{dt}\soc_{ij}\s\l(\bi,\bj\r)
-\nu\sum_{\bm}\D\s\l(\bm;\bi\r)\soc_{ij}\s\l(\bm,\bj\r)
+\sum_{\bm}\Big(\Wil\s\l(\bm,\bi\r)+\W\s\l(\bm;\bi\r)\delta_{il}\Big)
\soc_{lj}\s\l(\bm,\bj\r)
\notag\\
&
-\nu\sum_{\bm}\D\s\l(\bm;\bj\r)\soc_{ji}\s\l(\bm,\bi\r)
+\sum_{\bm}\Big(\Wjl\s\l(\bm,\bj\r)+\W\s\l(\bm;\bj\r)\delta_{jl}\Big)
\soc_{li}\s\l(\bm,\bi\r)
\notag\\
=&
-\sum_{\bm,\bn}\Al\s\l(\bm;\bn,\bi\r)\toc_{lij}\s\l(\bn,\bm,\bj\r)
-\,\sum_{\bm}\Bi\s\l(\bm;\bi\r)\overline{\tq\bbm\twj\bbj}
\notag\\ 
&
-\sum_{\bm,\bn}\Al\s\l(\bm;\bn,\bj\r)\toc_{lji}\s\l(\bn,\bm,\bi\r)
-\sum_{\bm}\Bj\s\l(\bm;\bj\r)\overline{\tq\bbm\twi\bbi},
\label{dtwitwjdt_Hom}
\end{align}
\begin{align}
-\sum_{\bm}\D\s\l(\bm;\bp\r)\overline{\tq\bbm\twi\bbi}=2\sum_{\bm}\Wk\s\l(\bm;\bp\r)\soc_{ki}\s\l(\bm,\bi\r)+
\sum_{\bm,\bn}\Clk\s\l(\bm,\bn;\bp\r)\toc_{kli}\s\l(\bm,\bn,\bi\r),
\label{dtqtwidt_Hom}
\end{align}
\begin{align}
&
\frac{d}{dt}\toc_{ijk}\s\l(\bi,\bj,\bk\r)
-\nu\sum_{\bm}\D\s\l(\bm;\bi\r)\toc_{ijk}\s\l(\bm,\bj,\bk\r)
+\sum_{\bm}\Big(\Wil\s\l(\bm,\bi\r)+\W\s\l(\bm;\bi\r)\delta_{il}\Big)
\toc_{ljk}\s\l(\bm,\bj,\bk\r)
\notag\\
&
-\nu\sum_{\bm}\D\s\l(\bm;\bj\r)\toc_{jki}\s\l(\bm,\bk,\bi\r)
+\sum_{\bm}\Big(\Wjl\s\l(\bm,\bj\r)+\W\s\l(\bm;\bj\r)\delta_{jl}\Big)
\toc_{lki}\s\l(\bm,\bk,\bi\r)
\notag\\ 
& 
-\nu\sum_{\bm}\D\s\l(\bm;\bk\r)\toc_{kij}\s\l(\bm,\bi,\bj\r)
+\sum_{\bm}\Big(\Wkl\s\l(\bm,\bk\r)+\W\s\l(\bm;\bk\r)\delta_{kl}\Big)
\toc_{lij}\s\l(\bm,\bi,\bj\r)
\notag\\
=&
-\sum_{\bm,\bn}\Al\s\l(\bm;\bn,\bi\r)
\foc_{lijk}\s\l(\bn,\bm,\bj,\bk\r)
-\sum_{\bm}\Bi\s\l(\bm;\bi\r)\overline{\tq\bbm\twj\bbj\twk\bbk}
\notag\\
&
-\sum_{\bm,\bn}\Al\s\l(\bm;\bn,\bj\r)
\foc_{ljki}\s\l(\bn,\bm,\bk,\bi\r)
-\sum_{\bm}\Bj\s\l(\bm;\bj\r)\overline{\tq\bbm\twk\bbk\twi\bbi}
\notag\\
&
-\sum_{\bm,\bn}\Al\s\l(\bm;\bn,\bk\r)
\foc_{lkij}\s\l(\bn,\bm,\bi,\bj\r)
-\sum_{\bm}\Bk\s\l(\bm;\bk\r)\overline{\tq\bbm\twi\bbi\twj\bbj}\,,
\label{dtwitwjtwkdt_Hom}
\end{align}
\begin{align}
-\sum_{\bl}\D\s\l(\bl;\bi\r)
\overline{\tq\bbl\twj\bbj\twk\bbk}
=
2\sum_{\bm}\Wl\s\l(\bm;\bi\r)\toc_{ljk}\s\l(\bm,\bj,\bk\r)
+\sum_{\bm,\bn}\Cli\s\l(\bm,\bn;\bi\r)
\foc_{iljk}\s\l(\bm,\bn,\bj,\bk\r),
\label{dtqtwitwjdt_Hom}
\end{align}
and
\begin{align}
-\sum_{\bl}\D\s\l(\bl;\bi\r)
\overline{\tq\bbl\tq\bbp}
=
2\sum_{\bm}\Wl\s\l(\bm;\bi\r)\overline{\tq\bbp\twl\bbm}
+\sum_{\bm,\bn}\Cjl\s\l(\bm,\bn;\bi\r)\overline{\tq\bbp\twl\bbm\twj\bbn}\,.
\label{dtqtqdt_Hom}
\end{align}

It is obvious that under the transformation of
\begin{align}
&
\Big\{
\soc_{ij}\s\l(\bi,\bj\r),\ 
\toc_{ijk}\s\l(\bi,\bj,\bk\r), \
\foc_{ijkl}\s\l(\bi,\bj,\bk,\bl\r), \
\overline{\tq\bbm\twi\bbi},\
\overline{\tq\bbl\twj\bbj\twk\bbk},\
\overline{\tq\bbl\tq\bbp}
\s\Big\} 
\notag\\
&\quad
\rightarrow
\Big\{
a\,\soc_{ij}\s\l(\bi,\bj\r), \ 
a\,\toc_{ijk}\s\l(\bi,\bj,\bk\r), \
a\,\foc_{ijkl}\s\l(\bi,\bj,\bk,\bl\r),\
a\,\overline{\tq\bbm\twi\bbi},\
a\,\overline{\tq\bbl\twj\bbj\twk\bbk},\
a\,\overline{\tq\bbl\tq\bbp}
\s\Big\}
\label{ScalingProperty}
\end{align}
where $a$ is any positive dimensionless constant, \eqref{dtwitwjdt_Hom} through \eqref{dtqtqdt_Hom} are invariant, along with all the constraints of equality and inequality discussed. This property of scaling invariance has certain implications for homogeneous turbulence modeling on the basis of objective optimization like $\K$ or $\Det$; Specifically, it rules out the possibility that, under a given $V_{ij}\not= 0$, a nontrivial bounded steady state solution is independent of the initial condition for the correlations. Otherwise, say, $\l\{\soc_{ij}^{\,\prime}\s\l(\bi,\bj\r)\r.$, $\toc_{ijk}^{\,\prime}\s\l(\bi,\bj,\bk\r)$, $\l.\foc_{ijkl}^{\,\prime}\s\l(\bi,\bj,\bk,\bl\r)\r\}$ were such a solution, then for any $a>1$, $\l\{a\,\soc_{ij}^{\,\prime}\s\l(\bi,\bj\r)\r.$, $a\,\toc_{ijk}^{\,\prime}\s\l(\bi,\bj,\bk\r)$, $\l.a\,\foc_{ijkl}^{\,\prime}\s\l(\bi,\bj,\bk,\bl\r)\r\}$ would be a solution too, and thus, the optimization scheme would produce an unbounded solution with the unboundedly great $\K$ and $\Det$.

We may infer from the above argument that solutions of the present homogeneous turbulence model for 
\begin{align*}
\l\{\soc_{ij}\s\l(\bi,\bj\r), \ 
\toc_{ijk}\s\l(\bi,\bj,\bk\r),\ 
\foc_{ijkl}\s\l(\bi,\bj,\bk,\bl\r),\ 
\overline{\tq\bbm\twi\bbi},\ \
\overline{\tq\bbl\twj\bbj\twk\bbk},\ \
\overline{\tq\bbl\tq\bbp}\r\}
\end{align*}
are time-dependent and initial condition dependent. This time-dependence appears compatible with the asymptotic behavior of homogeneous shear turbulence which shows the exponential growth of $\overline{\wk\wk}$ and so on, from the data of DNS and experiments (cf. \cite{SagautCambon2008}). To explore this issue further, we take 
\begin{align}
V_{ij}=S\,\delta_{i1}\delta_{j2},\ \ S> 0
\label{HomShear}
\end{align}
and introduce an asymptotic form solution of
\begin{align}
&
\soc_{ij}\s\l(\bi,\bj\r)=\soc_{ij}^{(0)}\s\l(\bi,\bj\r)\exp(\sigma S t),\quad 
\toc_{ijk}\s\l(\bi,\bj,\bk\r)=S\,\toc_{ijk}^{(0)}\s\l(\bi,\bj,\bk\r)\exp(\sigma S t),
\notag\\[5pt]
&
\foc_{ijkl}\s\l(\bi,\bj,\bk,\bl\r)=S^2\,\foc_{ijkl}^{(0)}\s\l(\bi,\bj,\bk,\bl\r)\exp(\sigma S t),\quad
\overline{\tq\bbm\twi\bbi}=S\,\overline{\tq\bbm\twi\bbi}^{(0)}\exp(\sigma S t),
\notag\\[5pt]
&
\overline{\tq\bbm\twi\bbi\twj\bbj}=S^2\,\overline{\tq\bbm\twi\bbi\twj\bbj}^{(0)}\exp(\sigma S t),\quad
\overline{\tq\bbl\tq\bbp}=S^2\,\overline{\tq\bbl\tq\bbp}^{(0)}\exp(\sigma S t)
\label{AsymptoticSolution}
\end{align}
where the quantities with superscript {\footnotesize $``(0)"$} are independent of time and $\sigma$ is a constant yet to be fixed. To see whether such a solution is admissible in the present model, we first substitute \eqref{HomShear} and \eqref{AsymptoticSolution} into \eqref{dtwitwjdt_Hom} through \eqref{dtqtqdt_Hom} to get
\begin{align}
&
\sigma \soc_{ij}^{(0)}\s\l(\bi,\bj\r)
-\frac{\nu}{S}\sum_{\bm}\D\s\l(\bm;\bi\r)\soc_{ij}^{(0)}\s\l(\bm,\bj\r)
+\sum_{\bm}\Big(\WilHOMS\s\l(\bm,\bi\r)+\WHOMS\s\l(\bm;\bi\r)\delta_{il}\Big)
\soc_{lj}^{(0)}\s\l(\bm,\bj\r)
\notag\\
&
-\frac{\nu}{S}\sum_{\bm}\D\s\l(\bm;\bj\r)\soc_{ji}^{(0)}\s\l(\bm,\bi\r)
+\sum_{\bm}\Big(\WjlHOMS\s\l(\bm,\bj\r)+\WHOMS\s\l(\bm;\bj\r)\delta_{jl}\Big)
\soc_{li}^{(0)}\s\l(\bm,\bi\r)
\notag\\
=&
-\sum_{\bm,\bn}\Al\s\l(\bm;\bn,\bi\r)\toc_{lij}^{(0)}\s\l(\bn,\bm,\bj\r)
-\sum_{\bm}\Bi\s\l(\bm;\bi\r)\overline{\tq\bbm\twj\bbj}^{(0)}
\notag\\ 
&
-\sum_{\bm,\bn}\Al\s\l(\bm;\bn,\bj\r)\toc_{lji}^{(0)}\s\l(\bn,\bm,\bi\r)
-\sum_{\bm}\Bj\s\l(\bm;\bj\r)\overline{\tq\bbm\twi\bbi}^{(0)},
\label{dtwitwjdt_HomShearAsympto}
\end{align}
\begin{align}
-\sum_{\bm}\D\s\l(\bm;\bp\r)\overline{\tq\bbm\twi\bbi}^{(0)}
=
\,2\sum_{\bm}\WkHOMS\s\l(\bm;\bp\r)\soc_{ki}^{(0)}\s\l(\bm,\bi\r)
+\sum_{\bm,\bn}\Clk\s\l(\bm,\bn;\bp\r)\toc_{kli}^{(0)}\s\l(\bm,\bn,\bi\r),
\label{dtqtwidt_HomShearAsympto}
\end{align}
\begin{align}
&
\sigma \toc_{ijk}^{(0)}\s\l(\bi,\bj,\bk\r)
-\frac{\nu}{S}\sum_{\bm}\D\s\l(\bm;\bi\r)\toc_{ijk}^{(0)}\s\l(\bm,\bj,\bk\r)
+\sum_{\bm}\Big(\WilHOMS\s\l(\bm,\bi\r)+\WHOMS\s\l(\bm;\bi\r)\delta_{il}\Big)
\toc_{ljk}^{(0)}\s\l(\bm,\bj,\bk\r)
\notag\\
&
-\frac{\nu}{S}\sum_{\bm}\D\s\l(\bm;\bj\r)\toc_{jki}^{(0)}\s\l(\bm,\bk,\bi\r)
+\sum_{\bm}\Big(\WjlHOMS\s\l(\bm,\bj\r)+\WHOMS\s\l(\bm;\bj\r)\delta_{jl}\Big)
\toc_{lki}^{(0)}\s\l(\bm,\bk,\bi\r)
\notag\\ 
& 
-\frac{\nu}{S}\sum_{\bm}\D\s\l(\bm;\bk\r)\toc_{kij}^{(0)}\s\l(\bm,\bi,\bj\r)
+\sum_{\bm}\Big(\WklHOMS\s\l(\bm,\bk\r)+\WHOMS\s\l(\bm;\bk\r)\delta_{kl}\Big)
\toc_{lij}^{(0)}\s\l(\bm,\bi,\bj\r)
\notag\\
=&
-\sum_{\bm,\bn}\Al\s\l(\bm;\bn,\bi\r)
\foc_{lijk}^{(0)}\s\l(\bn,\bm,\bj,\bk\r)
-\sum_{\bm}\Bi\s\l(\bm;\bi\r)\overline{\tq\bbm\twj\bbj\twk\bbk}^{(0)}
\notag\\
&
-\sum_{\bm,\bn}\Al\s\l(\bm;\bn,\bj\r)
\foc_{ljki}^{(0)}\s\l(\bn,\bm,\bk,\bi\r)
-\sum_{\bm}\Bj\s\l(\bm;\bj\r)\overline{\tq\bbm\twk\bbk\twi\bbi}^{(0)}
\notag\\
&
-\sum_{\bm,\bn}\Al\s\l(\bm;\bn,\bk\r)
\foc_{lkij}^{(0)}\s\l(\bn,\bm,\bi,\bj\r)
-\sum_{\bm}\Bk\s\l(\bm;\bk\r)\overline{\tq\bbm\twi\bbi\twj\bbj}^{(0)}\,,
\label{dtwitwjtwkdt_HomShearAsympto}
\end{align}
\begin{align}
-\sum_{\bl}\D\s\l(\bl;\bi\r)
\overline{\tq\bbl\twj\bbj\twk\bbk}^{(0)}
=
2\sum_{\bm}\WlHOMS\s\l(\bm;\bi\r)\toc_{ljk}^{(0)}\s\l(\bm,\bj,\bk\r)
+\sum_{\bm,\bn}\Cli\s\l(\bm,\bn;\bi\r)
\foc_{iljk}^{(0)}\s\l(\bm,\bn,\bj,\bk\r),
\label{dtqtwitwjdt_HomShearAsympto}
\end{align}
and
\begin{align}
-\sum_{\bl}\D\s\l(\bl;\bi\r)
\overline{\tq\bbl\tq\bbp}^{(0)}
=
2\sum_{\bm}\WlHOMS\s\l(\bm;\bi\r)\overline{\tq\bbp\twl\bbm}^{(0)}
+\sum_{\bm,\bn}\Cjl\s\l(\bm,\bn;\bi\r)\overline{\tq\bbp\twl\bbm\twj\bbn}^{(0)}\,.
\label{dtqtqdt_HomShearAsympto}
\end{align}
Here
\begin{align}
\WHOMS\s\l(\bm;\bk\r):=\frac{1}{S}\,\W\s\l(\bm;\bk\r),\quad
\WiHOMS\s\l(\bm;\bk\r):=\frac{1}{S}\,\Wi\s\l(\bm;\bk\r),\quad
\WijHOMS\s\l(\bm,\bk\r):=\frac{1}{S}\,\Wij\s\l(\bm,\bk\r)
\label{Coefficients_HomShearAsympto}
\end{align}
which are independent of $S$.
 As expected from the linear structure of \eqref{dtwitwjdt_Hom} through \eqref{dtqtqdt_Hom}, the algebraic equations above are independent of time. Next, we can verify directly that, under \eqref{AsymptoticSolution}, all the constraints of equality and inequality listed in Section~\ref{sec:BasicFormulation} and in \eqref{Constraints_HomogeneousTurbulence} can be converted into the corresponding constraints of the same form on $\soc_{ij}^{(0)}\s\l(\bi,\bj\r)$, $\toc_{ijk}^{(0)}\s\l(\bi,\bj,\bk\r)$, $\foc_{ijkl}^{(0)}\s\l(\bi,\bj,\bk,\bl\r)$, etc. without the presence of $S$. 

The values of $\sigma$, $\soc_{ij}^{(0)}\s\l(\bi,\bj\r)$, $\toc_{ijk}^{(0)}\s\l(\bi,\bj,\bk\r)$, $\foc_{ijkl}^{(0)}\s\l(\bi,\bj,\bk,\bl\r)$ and so on, depend on $\nu/S$ and $L$. To determine these values, we resort to maximize, say $\K=\sum_{\bk}\soc_{kk}\s\l(\bk,\bk\r)$; Due to the supposed constancy of $\sigma$ and the directional sequence of $t$, we will maximize
\begin{align}
\K=\sum_{\bk}\soc_{kk}\s\l(\bk,\bk\r)\Big|_{t\,=\,0}=\sum_{\bk}\soc_{kk}^{(0)}\s\l(\bk,\bk\r)
\label{ApproximateObjective_HomShearAsympto}
\end{align}
under \eqref{dtwitwjdt_HomShearAsympto} through \eqref{dtqtqdt_HomShearAsympto} and the resultant constraints of equality and inequality mentioned above. Next, within the peculiar context of the asymptotic form solution of \eqref{AsymptoticSolution} which holds mathematically for all $t$ $\in$ $(-\,\infty$, $+\infty)$, the flexibility of choosing any instant of time as the initial instant and the corresponding value change of the correlations obeying \eqref{AsymptoticSolution} motivates us to fix a specific upper bound for $\K$ above such as 
\begin{align}
\sum_{\bk}\soc_{kk}^{(0)}\s\l(\bk,\bk\r) \leq \K^{(0)}\quad\text{or}\quad
\sum_{\bk}\soc_{kk}^{(0)}\s\l(\bk,\bk\r) = \K^{(0)}.
\label{ApproximateObjective_HomShearAsympto_UpperBoundForInitialValue}
\end{align}
 Further detailed analysis and computation will check whether the solution matches with the data of DNS and experiments. Also, one can see the importance of developing an optimization algorithm and code so as to obtain the solution. 

One important issue yet to be resolved is whether, under a set of arbitrary initial conditions, a solution of \eqref{dtwitwjdt_Hom} through \eqref{dtqtqdt_Hom} exhibits the asymptotic behavior of \eqref{AsymptoticSolution}, without the explicit constraint of the latter.

As an apparently simple extension, one may also consider similarly the asymptotic solution under $V_{ij}$ $=$ $S \big(\delta_{i1}\delta_{j2}$ $+$ $\delta_{i2}\delta_{j1}\big)$ or  $V_{ij}$ $=$ $S \big(\delta_{i1}\delta_{j1}$ $-$ $\delta_{i2}\delta_{j2}\big)$. For more general homogeneous turbulence of $V_{ij}$, instead of \eqref{AsymptoticSolution}, we may seek an asymptotic solution of the form $\phi$ $=$ $\phi^{(0)}\exp(\sigma t)$.

\section{\label{sec:Remarks}Remarks}
\ \ \ \ We have discussed averaged turbulence modeling of multi-spatial scales for a turbulent flow of an incompressible Newtonian fluid in a bounded domain; the major points are summarized below.
\begin{enumerate}
  \renewcommand{\theenumi}{\roman{enumi}}
\item
The Reynolds average is supposedly adequate to partition the flow fields into the average ones and the corresponding turbulent fluctuations.
 There exists a function basis to decompose the fluctuations into the components associated with various spatial scales. Supposedly there is a probability density function $\pdf$ of these fluctuation components and there are well-defined statistical correlations of multi-scales among these fluctuation components (up to the fourth order). 
\item
The constraints of equality and inequality are formulated for the correlations in the wave number and physical spaces, which are either general intrinsic to turbulence or specific for the flow; the Cauchy-Schwarz inequality plays a significant role here. 
\item
The maximum information principle is resorted to determine the unbiased form of $\pdf$ under the equations of motion and the constraints of equality and inequality. Considering the formidable difficulty to implement the maximization of the information both analytical and computational, closely related but simple alternatives like the invariants of the covariance matrix $\big[\soc_{ij}(\bi,\bj)\big]$ are sought, two single objectives of special interest are, respectively, the determinant and the trace of the covariance matrix, the former suitable to the flow being slightly non-Gaussian and the latter due to its relation to the total fluctuation energy possessed in the turbulent flow.
\item
The model with the multi-scale correlations up to the fourth order is the smallest model capable of resolving the pressure fluctuation correlation, which does not contain additional coefficients besides the single objective to be maximized. Some preliminary results of its application to homogeneous turbulence are presented and the result on homogeneous shear turbulence suggests its possible compatibility with DNS and experimental data.
\end{enumerate}

To the goal of making the present formulation applicable, there are still many basic issues to be resolved, some of which are sketched below.
\begin{enumerate}
\item
It needs to be tested whether the information-based objective optimization reflects the physically observed phenomena. To check whether and how the model can produce meaningful results, we need to apply it to specific problems. Due to the number of variables and equations involved, it is preferable to consider some relatively simple problems first, like those of 1-D Burgers' equation and 2-D and 3-D homogeneous turbulence, with the help of various function bases. In this regard, the problem of homogeneous turbulence formulated in Section~\ref{sec:HomogeneousTurbulence} will be investigated further.
\item
From the viewpoint of computation and analysis, a reduced model containing the correlations only of the second order (and at most the third order) has a great advantage, which is also an extension of the Reynolds stress equation model of engineering turbulence. The task is then how to model adequately the third order correlation related terms on the right-hand side of \eqref{dtwitwjdt}, in the form  of  inequalities involving $\soc_{ij}\s\l(\bi,\bj\r)$ and $\toc_{ijk}\s\l(\bi,\bj,\bk\r)$, so as to make the reduced model determinate. \cite{TaoRamakrishnaRajagopal2008} attempted to address this issue, but the special tentative expression proposed is inadequate, for example, to satisfy the many constraints of equality and inequality involving the second order correlations listed in this work. 
\item
The treatments of representative wave numbers, locality, and local isotropy at high wave numbers are expected to reduce the number of correlations and equations involved in a model. Moreover, these treatments are motivated physically, and due to their physical nature, they supposedly help narrow the spread-out of $\pdf$ and make the predictions better. How to formulate the treatments is yet to be resolved, especially in the need to satisfy the constraints on the correlations, particularly those holding in the physical space.
\item
Some simple approximate single objective functions are presented, alternative to the information. Which approximate objective should be adopted is to be studied with the help of grid turbulence, etc.
\item
Pipe and channel flows of simple geometry will be simulated in order to test whether the model is adequate for wall-bounded turbulence.
\item 
An algorithm of optimization and solution is needed and essential which can deal with the great number of constraints.
\end{enumerate}

\section*{Acknowledgments}
\ \ \ \ L. Tao thanks IITM for its support of this work through the new faculty research grant (Grant number: ASE/06-07/187/NFSCLUOY).

\end{document}